\documentstyle[12pt]{article}
\def\Journal#1#2#3#4{{#1} {\bf #2}, #3 (#4)}

\def\NIMA{{\em Nucl.\ Instrum.\ Methods}\ A}
\def\NPA{{\em Nucl.\ Phys.}\ A}
\def\PLB{{\em Phys.\ Lett.}\  B}
\def\PRL{\em Phys.\ Rev.\ Lett.}
\def\PRD{{\em Phys.\ Rev.}\ D}
\def\PRC{{\em Phys.\ Rev.}\ C}
\def\ZPA{{\em Z.\ Phys.}\ A}
\begin{document}
\vspace*{-2cm}
\begin{center}
{\Large\bf Saturnalia\ \footnotetext{Concluding talk at ``les 20 ans de
Saturne-2'' colloquium, Paris, 5 May 1998}}\\[3ex]
{\large Colin Wilkin}\\[1ex]
{\normalsize University College London, London, WC1E 6BT, UK}\\[5ex]
\end{center}

\centerline{\bf Abstract}
\begin{verse}
{\it A personal choice is made of the highlights of the physics
programme carried out at the Saturne-2 facility over the last twenty years.}
\end{verse}

It is the end, the end of the meeting and the end of Saturne, and the kind
organisers have asked me to say a few words about both events.
We have been gathered here in Paris for the last two days to celebrate
the work of a whole collection of people and apparatus  grouped around an
accelerator Saturne-2, which closed down in December. I will not give
a political speech but there is an important point which has to be
made. The NSF facility at Daresbury was shut in the 1980's, at the height of
its productivity, by a committee of physicists who were faced with a 
budgetary crisis caused by the need to support particle
physics at a time of a rising Swiss Franc. When nuclear physicists
queried the action, the administration replied that it was a decision
taken on physics grounds, which is \underline{not} the same as a decision taken
physicists under pressure! As a result of this deliberate confusion,
which was
fed to the media, not only had the nuclear physicists lost their machine, 
but they felt devalued and insulted into the bargain. Do not let anyone tell 
you that Saturne could not have continued because there were not 
interesting problems still to be studied or that Saturne was not 
competitive for their study --- both would be false.

The flexibility of the accelerator, with its wide choice of particle types
and energy, and ease of energy changes, combined with an unrivalled collection
of spectrometers and polarimiters, made it a splendid tool for unravelling
problems in diverse branches of physics. Nevertheless the facility was closed.
Let us agree simply to say that the authorities had different priorities
to us.

Each of the excellent speakers has had to make a selection from the wealth of
material in the subject area of his talk and, in my half hour, I will
try to pick out some of the personal highlights of the whole programme of the
Laboratory and show how, in many cases, these have influenced
developments at other facilities. Though Saturne may be dead, its
genes live on!

Before starting this though, let me make one more
political point. It is much easier to get finance for doing new
experiments at other facilities, than completing the analysis of data
taken at Saturne. For the physicists involved it is also more
appealing to throw themselves wholeheartedly into the new
proposals. However we all know of many experiments, where there are
still data on tape representing interesting phenomena, but which risk
being left just gathering dust in cupboards. I will not embarass people
by giving examples but I appeal to physicists and their laboratory 
directors to spend some time and money to extract the maximum from the
data taken at Saturne.

It is exactly 30 years since I gave my first talk at Saclay, and in
May '68 people had other things on their minds than the application of
the Glauber model to the scattering of 1~GeV protons from carbon and
oxygen, which was the subject of the talk. Though the Saturne
accelerator was completely rebuilt 20 years ago, there has been a
continuity in its physics programme and in some of the equipment so
that I will stray across the 1977 frontier. Not being a
``machine-man'', there is of course a danger that I would not stress
sufficiently the role of equipment in my valediction. To avoid this
danger, I commend to you the excellent ``vulgarisation'' of the work
carried out at Saturne~\cite{Saturne}. Copies of the abridged English
translation were given to all participants, although they will be disappointed
not to find a description of the bonnes et mauvaises ann\'ees there!

One should also not forget the Proceedings of the Journ\'ees
d'\'Etudes Saturne, edited over many years by Pierre Radvanyi and
Mme.~Bordry. These volumes are mines of useful information about
Saturne, its equipment and its physics and this is a true reflection
of the meetings themselves, which were very important in developing
the unique culture of Saturne. As an example, my first ideas on the
possibility of quasi-bound $\eta\,^3$He states came to me at the Mont
Ste.\ Odile meeting, and the first calculations were carried out on the train
back from Strasbourg to Paris. Other people might remember rather the banquets
and the difference between Roscoff and Cavalaire!

Elastic scattering proton-nucleus scattering and the excitation of nuclear
levels was the first place where SPESI made an impact on the international
stage. As discussed here by Vorobyov, this programme was an extension of the
earlier Cosmotron work of Palevsky and friends. Vorobyov's talk reminded me
of the calculations of my youth, but I had truly forgotten that the
experimental programme of the Gatchina-Saclay colloaboration was so vast.
Nevertheless it was a programme of classical nuclear physics where the only
degrees of freedom were those of the nucleon. Within the Glauber model, they
obtained matter densities in many nuclei, and also transition densities,
without inserting specific forms for the nuclear densities. Thus they were
able to look at changes in density from one isobar to another.

Since Pierre Radvanyi and others have mentioned the influence of Palevsky
on the development of Saturne-1, allow me to add a couple of remarks to the
history since I worked with Harry in Brookhaven in 1966-67. Saturne-1 was a
clone of the Cosmotron, and Harry never understood why the French insisted
on taking the original plans for the machine rather than introducing the
improvements suggested by the American accelerator theorists who designed the
Cosmotron. Secondly, even though Palevsky is now recognised for his work on
elastic scattering, his real passion was for the comparison with the
inclusive spectra where he hoped to derive nucleon-nucleon correlations
from sum rules.

The first really
innovative material was an investigation of intermediate energy pick-up
reactions, where SPESI could easily
measure the angular distributions of $^{12}$C$(p,d)^{\,11}$C$^*$ to half a
dozen excited states of $^{11}$C at 700~MeV~\cite{Thirion}. It is fitting
that Jacques Thirion presented this at the same conference as Hoistad showed
data on the $(p,\pi^+)$ reaction, since the physics is closely related.
Unlike elastic proton scattering at small angles, these large momentum
transfer reactions are intimately related to mesonic or $\Delta$ degrees of
freedom in the nucleus. This importance of virtual mesons or nuclear isobars,
even for reactions where no mesons exist in either the initial or final
states, is something that we had to bear in mind in many later experiments
at Saturne-2. As a simple rule of thumb, rare events in hadronic reactions
tend not to be as rare
as would be indicated on the basis of purely nucleonic degrees of freedom.

Personally my most exciting moment at LNS was the night that we
measured $\eta$ production in the $\vec{d}p\to\, ^3$He$\,\eta$ reaction
near threshold at SPESIV~\cite{Berger}, though it had been studied
previously at higher energies~\cite{Berthet}. This was also an important 
moment for the laboratory as well since in this, and subsequent experiments 
at SPESII~\cite{Mayer1}:
\begin{itemize}
\item
It showed a surprisingly high cross section, which is probably due to a
virtual pion beam being created on one of the nucleons in the
deuteron, with the pion producing an $\eta$-meson on the second
nucleon. Such ideas are now used in the interpretation of threshold
meson production in ion-ion collisions.
\item
The variation of cross section near threshold implies the existence of
a quasibound $\eta\,^3$He state, a new form of nuclear matter at
high excitation energy, whose existence was clearly confirmed for the
$\eta\,^4$He system~\cite{Frascaria,Willis}.
\item
The latter provided a natural explanation for the charge symmetry violation
signal in the $dd\to\, ^4$He$\,\pi^0$ signal reported by the
unforgettable ER54 group just below the $\eta$ threshold~\cite{Goldzahl}. 
\item
It allowed the most precise measurement of the mass of the $\eta$
meson~\cite{Plouin1}.
\item
It opened the possibility of studying the rare decays of the $\eta$
with a tagged beam of such mesons~\cite{Mayer2}.
Important results were found on the decays $\eta\to\gamma\gamma$ and
$\eta\to \mu^+\mu^-$ and it is a matter of profound
regret that the SPESII programme of rare decay studies had to be cut short
for practical reasons. What a catastrophy.
\item
It lead to the study of heavier continuum states in
$pd\to\,^3$He$\,X$ under threshold conditions~\cite{Plouin2}, as well as the
isolation of heavier mesons $X$ such as the $\omega$~\cite{Ralf1}, $\eta'$ and
$\phi$~\cite{Ralf2}.
\item
It also drove people to investigate threshold production of $\eta$
\cite{Bergdolt,Pinot}, $\omega$~\cite{Hibou2}, $\eta'$~\cite{Hibou1} and 
$\phi$~\cite{Disto} in nucleon-nucleon scattering using a variety of 
experimental techniques.
\end{itemize}

Almost all of these themes, born out of one passionate night of endeavour, 
will be carried on at other experimental facilities, although the diehard 
might well say with some justification that we could have done it
still better at Saturne! Thus there is a
big rare decay programme at Uppsala and searches for quasibound
$\eta$-nuclear states at J\"{u}lich (and during the meeting Paul Kienle
explained to me that there were similar proposals at GSI). Both laboratories
will carry out production studies of the $\eta^{\,3}$He system and search
for charge symmetry violation in $dd\to ^{\,4}$He$\,\pi^0$. I have even
heard it said that by 2003 Fran\c{c}ois Plouin will have finished his analysis
 of the mass of the $\eta$ meson, but I have no confirmation of this.

Saturne was internationally recognised for the intensity and quality
of its beams of polarised protons and deuterons~\cite{Saturne}, which 
was due to the combination of the Hyperion polarised ion source,
the mini-synchrotron Mimas serving as injector, and a meticulous study
of depolarising resonances in Saturne itself by the ``machinistes''. 
I was intitially confused to find that the ``Group Th\'eorie'' at
Saturne was composed purely of people who studied how the machine worked,
whereas there were almost no nuclear/particle theorists in the laboratory. 
This showed
the priorities of the founders of the laboratory and all other
machines with which I am familiar have far less backing in this area. 
Though the
successes in Indiana have been outstanding, it has been found that polarised
particles are difficult to accelerate at both Uppsala and J\"ulich. In his
talk Lagniel highlighted the collaborations between machinistes and
experimentalists, especially those of nucleon-nucleon, and showed how both
sides benefitted from this. It wasn't always war!

The major user of polarised protons was the nucleon-nucleon group and often
in the Comit\'e des Exp\'eriences we exploited the fact that there was always
a colleague who was willing to take beam time during any holiday period. 
It might be a coincidence but the moment that they finished their programme
on proton-proton and proton-neutron elastic scattering, the Authorities
closed the Laboratory down. The nucleon-nucleon studies represented perhaps
the most fundamental of the research that was carried out at Saturne. This was
done in a completely professional manner and in many cases the group had
sufficient data in order to make direct reconstructions of the amplitudes
without passing through theoretical models.

Where Saturne was really unique was in respect of its deuteron beams,
which combined high intensity with large values of both vector \underline{and}
tensor polarisation. Without Saturne we could not have calibrated the
AHEAD~\cite{AHEAD} or POLDER~\cite{POLDER} deuteron tensor polarimeters 
which, as Gar\c{c}on described in his talk, played vital roles in
separating the deuteron form factors through measurements of the
polarisation of the recoil deuteron in elastic electron-deuteron
scattering. The Comit\'e des Exp\'eriences and Management of Saturne always 
took the broad view that the advancement of Physics was the central theme
of the Laboratory and, if this involved calibrating equipment for 
\underline{good} experiments to be done elsewhere, then this was a
price worth paying. It certainly paid off in the case of the deuteron
form factor. Michel Gar\c{c}on has shown us the first preliminary results
from CEBAF on the value of T$_{20}$ in elastic $ed\to e'\vec{d}'$. The
collaboration could work to values of $q^2$ twice as big as earlier
experiments with a much bigger efficiency than all competing techniques.

The POLDER polarimeter is based upon the deuteron charge exchange
reaction $\vec{d}p \to \{pp\}n$, which was predicted to show a strong 
analysing power signal if the final proton-proton pair had small
excitation energy and was produced with a small momentum transfer
relative to the incident deuteron~\cite{BW}. This reaction was
investigated in the few hundred MeV range~\cite{EMRIC} but also in the
GeV range through the remarkable technique of detecting both protons
in the focal plane of the SPESIV spectrometer~\cite{Gaarde1}. This was
one of a whole series of charge-exchange experiments with different probes
undertaken over years by the Copenhagen-Orsay group, and it is a matter of
profound regret that the group leader, Carl Gaarde, was taken from us a few
weeks before this meeting. Michelle Roy-St\'ephan has reminded us of the
enormous quantity of results, especially in the ($^3$He,t) reaction,
achieved over 14 years and the successes in their interpretation
has come through active collaborations with theorists. One of these,
Madeleine Soyeur, showed how such results could be exciting for theorists
and I believe that she communicated well this enthusiasm to the machinistes
in the hall, even if they couldn't appreciate some of the finer points.

The method of detecting two protons in SPESIV could be extended to looking
also at deep inelastic scattering, where a $\Delta$-isobar had been produced
on hydrogen or nuclear targets~\cite{Gaarde2}. The $(\vec{d},2p)$ reaction
is an interesting probe for $\Delta T = \Delta S =1$ transitions but,
inspired by Marcel Morlet, the Orsay group
showed that measurements of vector spin transfer in 
$(\vec{d},\,\vec{d}^{\,'})$ can provide useful signals for $\Delta T=0$,  
$\Delta S =1$ states in the residual nucleus~\cite{Morlet}. This
needed a lot of intuition because the symmetry used here is only an
approximate one and a complete separation would have required the
combination of a tensor polarised beam and tensor polarimeter.

Of equally lasting importance was the whole series of measurements
designed to investigate the few body problem in intermediate energy
nuclear physics. The measurement of triple spin parameters in 
$\vec{d}\vec{p}\to\vec{p}d$ at 1.6~GeV~\cite{Igo} represented a real 
tour de force which is never likely to be repeated. More appealing are
however the survey experiments which measure a single observable, such
as the deuteron tensor analysing power T$_{20}$, as a function of beam
energy, as was done for $\vec{d}p \to pd$ in the backward direction 
\cite{Arvieux} or pion production through $\vec{d}p\to\,^{3}$He$\,\pi^0$
\cite{Kerboul,Nikulin}. Due to its D-state component, the deuteron has
a geometric deformation looking in momentum space a bit like a
pancake. This could be investigated by measuring the T$_{20}$
dependence of the Fermi momentum in deuteron break up $\vec{d}p \to
ppn$~\cite{Punjabi1} but the wealth of polarimeters at Saturne also
permitted a polarisation measurement of a final proton~\cite{Stan}.

The D-states in $^6$Li could be studied in analogous break-up
measurements~\cite{Punjabi2} with beams of polarised ions of several 
GeV/c (which must be high compared to the Fermi momentum), and this seems
to have been the only \underline{published} experiment using such a beam.
The MIMAS synchrotron was designed to furnish high quality polarised
beams and also beams of ions up to $^{129}$Xe (30$^{+}$) but the demand 
for such heavy ions has represented only about 10\% of the total 
requests over many years, and even less of the publications. In part this
might be that
the competion from other facilities in the heavy ion field, such as Ganil 
or SIS, has proved more intense than for the light ions where Saturne was
supreme. However, apart from Diog\`ene, there were no specific heavy ion 
detectors at Saturne and groups tended to bring small equipment with
them for multifragmentation or other studies~\cite{Gelbke}. Saturne
was the great success that it was for light ions because it had a whole
collection of outstanding spectrometers which were tailored for the purpose. 

I have already stressed that physics must be exciting and
no survey of the work at Saturne would be complete without mentioning
``Le pari de Pascal'' and the human desire to find something really
revolutionary, even if the probability of success were infinitesimally
small. If there were a good bridge player in the room, then he would 
tell you that you must think defensively, such that the international 
competitors should never be allowed to make too good a score. Thus Saturne 
could not allow other laboratories to make the earth-shattering 
discoveries which might have been made at Saclay. In other fields this
maladie took the form of trying to repeat the discoveries of Pons and 
Fleischmann on cold fusion, but at Saturne it was the hunt for 
narrow dibaryon resonances~\cite{Boris1}, pion-nucleon bound states 
\cite{Boris2}, abnormal production of pions~\cite{Julien}, the
anomalons~\cite{anomalon} {\it etc.}
It was really like something out of ``Les vacances de M.~Hulot''. 
Nothing convincing was ever found which was not at the limits of 
systematic/statistical error bars. Colleagues will be relieved to know
that this approach is being carried on in the successor laboratories
and that I was equally unsuccessful in stopping the search for the $d'$
at COSY and CELSIUS PAC's as I was with the narrow dibaryons at the
Saturne Comit\'e des Exp\'eriences. 

It is true that just one such discovery might have provided some
defence for Saturne and it is often these searches which bring out the best in
technical ingenuity. One example of this is the famous wheel of
Beurtey and Saudinos~\cite{Beurtey}. In order to make simultaneous
measurements of $A_y$ in proton-proton elastic scattering at many
different energies, they constructed an energy degrader with many
steps, looking something like a circular Escher staircase. By letting
this rotate fast in the beam, high statistics data could be obtained
at 16 different energies simultaneously. A short run was sufficient to
kill off one dibaryon (but not French).

Over the years Saturne has been invaluable for the calibration of much
other equipment in addition to polarimeters. It is true that Saturne will
\underline{not} be much remembered for such mundane work in the publications 
of physics carried out some time later at other laboratories. 
However people will remember
Saturne when they find that they can no longer use it to test counters for
LHC for example. Often the applied research at Saturne gave rise to
interesting physics in related fields, and I have in mind here the some of
the work described by Rolf Michel on pseudo-meteorites. 

You will have noticed that I have not spoken at all about the three big
programmes which have dominated the last one or two years of the running
of Saturne-2, {\it viz} DISTO, SPESIV$\pi$, and Nuclear Transmutation;
it is much too early to judge their significance. We all of course hope
that the information provided by the transmutation experiments will prove
valuable in a search for a sensible method for dealing with nuclear waste.
That really would be a worthwhile monument for Saturne to compete with the
statue of the Ptolemy which was unveiled today outside the Grand Palais.\\

Pour terminer, je dois remercier Alain, Fran\c{c}oise, Simone et Bernard
qui ont assur\'e le succ\`es de cette r\'eunion.
Il est bien \'evident que le fran\c{c}ais n'est pas la langue
maternelle, ni de M.~Blair ni de moi. N\'eanmoins il va rester ma
langue fraternelle et cela est d'une grande part d\^u aux
collaborations fructueuses avec les gens autour de Saturne, beaucoup
tr\^op nombreux \`a mentionner.

Saturne, nous allons tous nous souvenir de vous. Saturne, adieu! Merci!

\end{document}